\documentclass[multphys,vecphys]{svmult}

\usepackage{makeidx}         
\usepackage{graphicx}        
\usepackage{multicol}        
\usepackage[bottom]{footmisc}

\makeindex             

\begin{document}

\title*{Dynamics and Shape of Brightest Cluster Galaxies}
\author{H.\ Andernach\inst{1}, K.\ Alamo-Mart\'{\i}nez\inst{1},
R.\ Coziol\inst{1} and E.\ Tago\inst{2}}
\institute{Departamento de Astronom\'{\i}a, Universidad de Guanajuato, Mexico
\texttt{heinz@astro.ugto.mx}, \texttt{karla@astro.ugto.mx}, \texttt{rcoziol@astro.ugto.mx}
\and Tartu Observatory, T\~oravere, Estonia~~
\texttt{erik@aai.ee}}
%
%
\maketitle

\begin{abstract}

We identified Brightest Cluster Members (BCM) on DSS images of 1083 Abell 
clusters, derived their individual and host cluster redshifts
from literature and determined the BCM ellipticity. Half the BCMs move
at a speed higher than 37\% of the cluster velocity dispersion $\sigma_{\rm cl}$,
suggesting that most BCMs are part of substructures falling into the main
cluster. Both, the BCM's velocity offset in units of $\sigma_{\rm cl}$,
and BCM ellipticity, weakly decrease with cluster richness.
\end{abstract}


If BCMs formed ``in-situ'' they should ``rest'' at the bottom 
of the potential well of the cluster and have a radial velocity
equal to the cluster mean. Previous studies, e.g.\ \cite{Zabludoffetal1993}, 
based on small samples of clusters dominated by a cD galaxy, showed 
that this is far from reality.
Here we extend these studies to 1083 Abell clusters likely to have a dominant
galaxy, namely all clusters with Bautz-Morgan (BM) types I or I-II,
all with Rood-Sastry (RS) type cD, and all clusters with notes in \cite{aco89}
indicating a ``corona'' of the first-ranked galaxy. From DSS images
we derived positions for 1329 BCM candidates in 1076 clusters
(occasionally more than one candidate per cluster), retrieved their basic 
parameters from NED, and extracted cluster mean redshifts, $\rm{z}_{cl}$, 
and velocity dispersions, $\sigma_{\rm cl}$, from the compilation maintained by two of us 
\cite{Andernachetal2005}. Deleting all foreground/background BCM candidates
and restriction to clusters with $\ge$10 measured redshifts, yielded a sample 
of 385 BCMs in 326 Abell clusters for which we derived the
relative velocity offset, 
$\rm{v}_{off}/\sigma_{cl} = (\rm{v}_{BCM} - c\rm{z}_{cl})/(1+z_{\rm cl})/\sigma_{cl}$,
with v$_{\rm BCM}$ the BCM radial velocity.

\begin{figure}
\centering
\includegraphics[height=63mm]{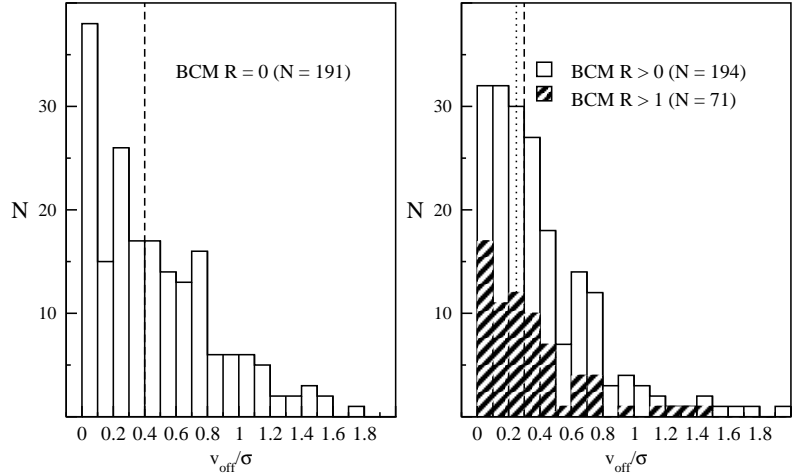}
%
%
\caption{Relative velocity offset for clusters with different Abell richness~R. 
Median values are indicated with dashed lines or a dotted line for 
the shaded histogram.}
\label{and2:fig1}       
\end{figure}

In Fig.~\ref{and2:fig1} we show that half of the BCMs in our sample 
move at peculiar velocities $> 0.37\,\sigma_{\rm cl}$. There is a trend for 
a smaller $\rm{v}_{off}/\sigma_{cl}$ in richer clusters which is expected if 
the latter are dynamically more evolved.

\begin{figure}
\centering
\includegraphics[height=63mm]{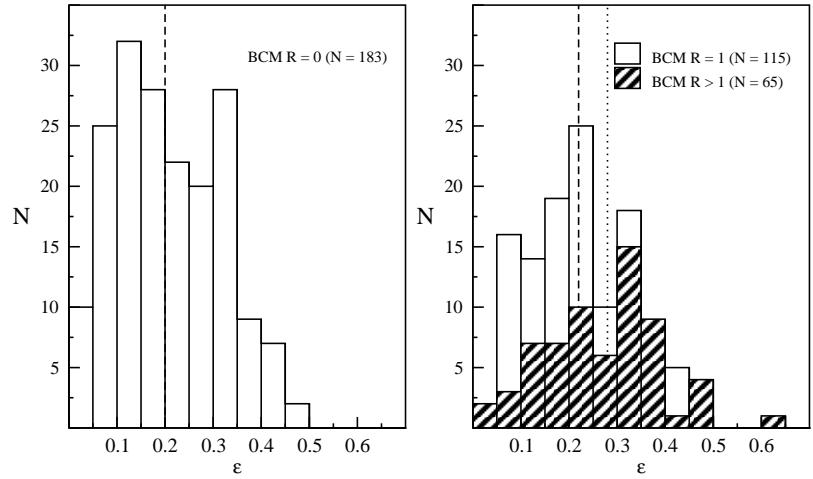}
%
%
\caption{BCM ellipticity distribution in clusters of different Abell richness. 
Median values are indicated with dashed
lines or a dotted line for the shaded histogram. }
\label{and2:fig2}       
\end{figure}

We also determined BCM ellipticities using IRAF's {\tt ellipse} for 1193
BCM candidates in 1012 clusters, but restricted our statistical analysis
to the same sample of 385 BCMs mentioned above. Figure~\ref{and2:fig2} shows
that BCM ellipticity increases with richness.  This may suggest that
BCMs in richer clusters grow more likely by anisotropic mergers.

Our observations are consistent with a model where most BCM form
during the collapse and virialization of poor clusters or compact
groups with low velocity dispersions \cite{Merritt1985}. This
supports the view that most galaxies formed in groups (and not in
rich clusters) with a common dark halo and/or individual halo of
each galaxy which form(s) a local potential minimum for the BCM.

%
%
%

%
%



\printindex
\end{document}